\documentclass[a4paper]{report}
\usepackage[utf8]{inputenc}
\usepackage[T1]{fontenc}
\usepackage{RJournal}
\usepackage{amsmath,amssymb,array}
\usepackage{booktabs}

\providecommand{\tightlist}{%
  \setlength{\itemsep}{0pt}\setlength{\parskip}{0pt}}

\begin{document}

\sectionhead{Contributed research article}
\volume{XX}
\volnumber{YY}
\year{20ZZ}
\month{AAAA}

\begin{article}
\title{A Simple Guide to S3 Methods}
\author{by Nicholas Tierney}

\maketitle

\abstract{%
Writing functions in R is an important skill for anyone using R. S3
methods allow for functions to be generalised across different classes
and are easy to implement. Whilst many R users are be adept at creating
their own functions, it seems that there is room for many more to take
advantage of R's S3 methods. This paper provides a simple and targeted
guide to explain what S3 methods are, why people should them, and how
they can do it.
}

\section{Introduction}\label{introduction}

A standard principle of programming is DRY - Don't Repeat Yourself.
Under this axiom, the copying and pasting of the same or similar code
(copypasta), is avoided and instead replaced with a function, macro, or
similar. Having one function to replace several of the same or similar
coded sections simplifies code maintenance as it means that only one
section of code needs to be maintained, instead of several. This means
that if the code breaks, then one simply needs to update the function,
rather than finding all of the coded sections that are now broken.

S3 methods in the R programming language are a way of writing functions
in R that do different things for objects of different classes. S3
methods are so named as the methods shipped with the release of the
third version of the ``S'' programming language, which R was heavily
based upon (Chambers and Hastie 1992, Team (2016), R Core Team (2012)).
Hence, methods for S 3.0 = S3 Methods.

The function \texttt{summary()} is an S3 method. When applied to an
object of class \texttt{data.frame}, \texttt{summary} shows descriptive
statistics (Mean, SD, etc.) for each variable. For example,
\texttt{iris} is of class \texttt{data.frame}:

\begin{Schunk}
\begin{Sinput}
class(iris)
\end{Sinput}
\begin{Soutput}
#> [1] "data.frame"
\end{Soutput}
\end{Schunk}

So applying \texttt{summary} to \texttt{iris} gives us summary
information relevant to a dataframe

\begin{Schunk}
\begin{Sinput}
summary(iris)
\end{Sinput}
\begin{Soutput}
#>   Sepal.Length    Sepal.Width     Petal.Length    Petal.Width   
#>  Min.   :4.300   Min.   :2.000   Min.   :1.000   Min.   :0.100  
#>  1st Qu.:5.100   1st Qu.:2.800   1st Qu.:1.600   1st Qu.:0.300  
#>  Median :5.800   Median :3.000   Median :4.350   Median :1.300  
#>  Mean   :5.843   Mean   :3.057   Mean   :3.758   Mean   :1.199  
#>  3rd Qu.:6.400   3rd Qu.:3.300   3rd Qu.:5.100   3rd Qu.:1.800  
#>  Max.   :7.900   Max.   :4.400   Max.   :6.900   Max.   :2.500  
#>        Species  
#>  setosa    :50  
#>  versicolor:50  
#>  virginica :50  
#>                 
#>                 
#> 
\end{Soutput}
\end{Schunk}

\texttt{summary} also performs differently when applied to different
object. In fact, you can find all the classes that work with an S3
method by typing the following:

\begin{Schunk}
\begin{Sinput}
methods(summary)
\end{Sinput}
\begin{Soutput}
#>  [1] summary.aov                    summary.aovlist*              
#>  [3] summary.aspell*                summary.check_packages_in_dir*
#>  [5] summary.connection             summary.data.frame            
#>  [7] summary.Date                   summary.default               
#>  [9] summary.ecdf*                  summary.factor                
#> [11] summary.glm                    summary.infl*                 
#> [13] summary.lm                     summary.loess*                
#> [15] summary.manova                 summary.matrix                
#> [17] summary.mlm*                   summary.nls*                  
#> [19] summary.packageStatus*         summary.PDF_Dictionary*       
#> [21] summary.PDF_Stream*            summary.POSIXct               
#> [23] summary.POSIXlt                summary.ppr*                  
#> [25] summary.prcomp*                summary.princomp*             
#> [27] summary.proc_time              summary.srcfile               
#> [29] summary.srcref                 summary.stepfun               
#> [31] summary.stl*                   summary.table                 
#> [33] summary.tukeysmooth*          
#> see '?methods' for accessing help and source code
\end{Soutput}
\end{Schunk}

There's over 30 different methods!

We can use summary on a linear model, for example:

\begin{Schunk}
\begin{Sinput}
lm_iris <- lm(Sepal.Length ~ Sepal.Width, data = iris)

summary(lm_iris)
\end{Sinput}
\begin{Soutput}
#> 
#> Call:
#> lm(formula = Sepal.Length ~ Sepal.Width, data = iris)
#> 
#> Residuals:
#>     Min      1Q  Median      3Q     Max 
#> -1.5561 -0.6333 -0.1120  0.5579  2.2226 
#> 
#> Coefficients:
#>             Estimate Std. Error t value Pr(>|t|)    
#> (Intercept)   6.5262     0.4789   13.63   <2e-16 ***
#> Sepal.Width  -0.2234     0.1551   -1.44    0.152    
#> ---
#> Signif. codes:  0 '***' 0.001 '**' 0.01 '*' 0.05 '.' 0.1 ' ' 1
#> 
#> Residual standard error: 0.8251 on 148 degrees of freedom
#> Multiple R-squared:  0.01382,    Adjusted R-squared:  0.007159 
#> F-statistic: 2.074 on 1 and 148 DF,  p-value: 0.1519
\end{Soutput}
\end{Schunk}

\texttt{summary} produces a description of the linear model, describing
how it was called (\texttt{call}), as well as the \texttt{residuals},
\texttt{coefficients}, \texttt{t-values}, \texttt{p-values}, \(R^2\),
and more. This output is \textbf{completely} different to the
information output from \texttt{summary} used for the \texttt{iris}
dataframe.

So how does the same function, \texttt{summary} perform differently for
different objects? The answer is that R is helpful, and \emph{hides}
this information. There are in fact, many different \texttt{summary}
functions. For example:

\begin{itemize}
\tightlist
\item
  \texttt{summary.lm}
\item
  \texttt{summary.data.frame}
\item
  \texttt{summary.Date}
\item
  \texttt{summary.matrix}
\end{itemize}

Being an S3 method, \texttt{summary} calls the appropriate function
based upon the class of the object it operates on. So using
\texttt{summary} on an object of class ``Date'' will evoke the function,
\texttt{summary.Date}. \textbf{But all you need to do is type
\texttt{summary}}, and the S3 method does the rest. By abstracting away
this detail (the object class), the intent becomes clearer.

To further illustrate, using \texttt{summary} on the \texttt{iris} data
will actually call the function \texttt{summary.data.frame}, since
\texttt{iris} is of class \texttt{data.frame}. We can find the class of
an object using \texttt{class}

\begin{Schunk}
\begin{Sinput}
class(iris)
\end{Sinput}
\begin{Soutput}
#> [1] "data.frame"
\end{Soutput}
\end{Schunk}

\begin{Schunk}
\begin{Sinput}
summary.data.frame(iris)
\end{Sinput}
\begin{Soutput}
#>   Sepal.Length    Sepal.Width     Petal.Length    Petal.Width   
#>  Min.   :4.300   Min.   :2.000   Min.   :1.000   Min.   :0.100  
#>  1st Qu.:5.100   1st Qu.:2.800   1st Qu.:1.600   1st Qu.:0.300  
#>  Median :5.800   Median :3.000   Median :4.350   Median :1.300  
#>  Mean   :5.843   Mean   :3.057   Mean   :3.758   Mean   :1.199  
#>  3rd Qu.:6.400   3rd Qu.:3.300   3rd Qu.:5.100   3rd Qu.:1.800  
#>  Max.   :7.900   Max.   :4.400   Max.   :6.900   Max.   :2.500  
#>        Species  
#>  setosa    :50  
#>  versicolor:50  
#>  virginica :50  
#>                 
#>                 
#> 
\end{Soutput}
\end{Schunk}

which is the same as \texttt{summary(iris)}

\begin{Schunk}
\begin{Sinput}
sum1_df <- summary.data.frame(iris)

sum2_df <- summary(iris)

all.equal(sum1_df, sum2_df)
\end{Sinput}
\begin{Soutput}
#> [1] TRUE
\end{Soutput}
\end{Schunk}

And using summary on the linear model object, \texttt{lm\_iris}
performs:

\begin{Schunk}
\begin{Sinput}
summary.lm(lm_iris)
\end{Sinput}
\begin{Soutput}
#> 
#> Call:
#> lm(formula = Sepal.Length ~ Sepal.Width, data = iris)
#> 
#> Residuals:
#>     Min      1Q  Median      3Q     Max 
#> -1.5561 -0.6333 -0.1120  0.5579  2.2226 
#> 
#> Coefficients:
#>             Estimate Std. Error t value Pr(>|t|)    
#> (Intercept)   6.5262     0.4789   13.63   <2e-16 ***
#> Sepal.Width  -0.2234     0.1551   -1.44    0.152    
#> ---
#> Signif. codes:  0 '***' 0.001 '**' 0.01 '*' 0.05 '.' 0.1 ' ' 1
#> 
#> Residual standard error: 0.8251 on 148 degrees of freedom
#> Multiple R-squared:  0.01382,    Adjusted R-squared:  0.007159 
#> F-statistic: 2.074 on 1 and 148 DF,  p-value: 0.1519
\end{Soutput}
\end{Schunk}

the same as \texttt{summary(lm\_iris)}

\begin{Schunk}
\begin{Sinput}
sum1_lm <- summary.lm(lm_iris)

sum2_lm <- summary(lm_iris)

all.equal(sum1_lm, sum2_lm)
\end{Sinput}
\begin{Soutput}
#> [1] TRUE
\end{Soutput}
\end{Schunk}

One could coerce a different method upon a different class, for example
using \texttt{summary.data.frame} on an ``lm'' object:

\begin{Schunk}
\begin{Sinput}
summary.data.frame(lm_iris)
\end{Sinput}
\begin{Soutput}
#>   coefficients       residuals          effects               rank  
#>  Min.   :-0.2234   Min.   :-1.5561   Min.   :-71.56593   Min.   :2  
#>  1st Qu.: 1.4640   1st Qu.:-0.6333   1st Qu.: -0.65192   1st Qu.:2  
#>  Median : 3.1514   Median :-0.1120   Median : -0.00897   Median :2  
#>  Mean   : 3.1514   Mean   : 0.0000   Mean   : -0.42040   Mean   :2  
#>  3rd Qu.: 4.8388   3rd Qu.: 0.5579   3rd Qu.:  0.61051   3rd Qu.:2  
#>  Max.   : 6.5262   Max.   : 2.2225   Max.   :  2.15225   Max.   :2  
#>  fitted.values       assign     qr.Length  qr.Class  qr.Mode  df.residual 
#>  Min.   :5.543   Min.   :0.00   300      -none-   numeric    Min.   :148  
#>  1st Qu.:5.789   1st Qu.:0.25     2      -none-   numeric    1st Qu.:148  
#>  Median :5.856   Median :0.50     2      -none-   numeric    Median :148  
#>  Mean   :5.843   Mean   :0.50     1      -none-   numeric    Mean   :148  
#>  3rd Qu.:5.901   3rd Qu.:0.75     1      -none-   numeric    3rd Qu.:148  
#>  Max.   :6.080   Max.   :1.00                                Max.   :148  
#>    xlevels         call         terms        
#>  Length:0      Length:3      Length:3        
#>  Class :list   Class :call   Class1:terms    
#>  Mode  :list   Mode  :call   Class2:formula  
#>                              Mode  :call     
#>                                              
#>                                              
#>  model.Sepal.Length  model.Sepal.Width 
#>  Min.   :4.300000    Min.   :2.000000  
#>  1st Qu.:5.100000    1st Qu.:2.800000  
#>  Median :5.800000    Median :3.000000  
#>  Mean   :5.843333    Mean   :3.057333  
#>  3rd Qu.:6.400000    3rd Qu.:3.300000  
#>  Max.   :7.900000    Max.   :4.400000
\end{Soutput}
\end{Schunk}

However the output may be a bit confusing.

To summarize, the important feature of S3 methods worth noting is that
only the \textbf{first part}, \texttt{summary}, is required to be used
on these objects of different classes.

\section{Why hide the text?}\label{why-hide-the-text}

Hiding the trailing text after the \texttt{.} avoids the need to use a
different \texttt{summary} function for every class. This means that one
does not need to remember to use \texttt{summary.lm} for linear models,
or \texttt{summary.data.frame} for data frames, or
\texttt{summary.aProposterousClassOfObject}. By using S3 methods,
cognitive load is reduced - you don't have to think as much to remember
what class an object is - and the commands are more intuitive. To get a
summary of most objects, use \texttt{summary}, to plot most objects, use
\texttt{plot}. Perhaps the most nifty feature of all is that a user can
create their own S3 methods using the same functions such as
\texttt{summary} and \texttt{plot}. This means a user can create their
own special class of object

\begin{Schunk}
\begin{Sinput}
test_class <- 1:10

class(test_class) <- "myclass"

class(test_class)
\end{Sinput}
\begin{Soutput}
#> [1] "myclass"
\end{Soutput}
\end{Schunk}

and then write their own S3 method for it - e.g.,
\texttt{summary.myclass} or \texttt{plot.myclass}, each proiding
appropriate summary information, or nice plots, for that object.

\section{How to make your own S3
method?}\label{how-to-make-your-own-s3-method}

Creating your own S3 method is not particularly difficult and is usually
highly practical. A use case scenario for creating an S3 method is now
discussed.

The Residual Sums of Squares (RSS), \(\sum(Y_i - \hat{Y})^2\) is a
useful metric for determining model accuracy for continuous outcomes.
For example, for a Classification and Regression Tree

\begin{Schunk}
\begin{Sinput}
library(rpart)

fit.rpart <- rpart(Sepal.Width ~ Sepal.Length + Petal.Length + Petal.Width + Species, data = iris)
\end{Sinput}
\end{Schunk}

The RSS is calculated as

\begin{Schunk}
\begin{Sinput}
print_rss <- sum(residuals(fit.rpart)**2)
\end{Sinput}
\end{Schunk}

One might be inclined to write a function to perform this task

\begin{Schunk}
\begin{Sinput}
rss <- function(x){
  
  sum(residuals(x)**2)
  
}

rss(fit.rpart)
\end{Sinput}
\begin{Soutput}
#> [1] 10.17245
\end{Soutput}
\end{Schunk}

However, there are many different decision tree models that one would
like to compare, say boosted regression trees (BRT), and random forests
(RF). The same code will not work:

\begin{Schunk}
\begin{Sinput}
library(randomForest)
set.seed(71)
fit.rf <- randomForest(Sepal.Length ~ ., data=iris, importance=TRUE,
                        proximity=TRUE)

rss(fit.rf)
\end{Sinput}
\begin{Soutput}
#> [1] 0
\end{Soutput}
\end{Schunk}

In this case, one could write three functions, one for each decision
tree method: ``rss\_rpart'', ``rss\_brt'', and ``rss\_rf''. But to avoid
having three functions and instead use just one, one could place all
three functions inside of one function, using an if-then-else clause to
direct the object of the appropriate class to the appropriate method.
This shall be referred to as a ``Poor man's S3 method''.

\begin{verbatim}
dt_rss <- function (x){


if ("rpart" %in% class(x)) {
  
  result <- sum((residuals(x)**2))
  
  return(result)
  
}

else if ("gbm" %in% class(x)) {
  
  result <- sum(x$residuals**n2)
  
  return(result)

}

else if ("randomForest" %in% class(x)) {

  temp <- x$y - x$predicted  
  
  result <- sum(temp**2)  
  
  return(result)

}
  
else warning(paste(class(x), "is not of an rpart, gbm, or randomForest object"))
}
\end{verbatim}

Here it is in action

\begin{verbatim}
dt_rss(fit.rpart)

#> [1] 10.17245
\end{verbatim}

The RSS method works, and if it is applied to a class that is not known,
a special message is provided

\begin{Schunk}
\begin{Sinput}
fit.lm <- lm(Sepal.Width ~ Species, data = iris)

dt_rss(fit.lm)
\end{Sinput}
\begin{Soutput}
#> Warning in dt_rss(fit.lm): lm is not of an rpart, gbm, or randomForest
#> object
\end{Soutput}
\end{Schunk}

The ``poor man's S3 method'' does what it needs to do. However, one must
ask how sustainable this would be for an entire programming language?
Imagine if a colleague creates a new tree method that needs it's own
\texttt{rss()}. He will need to convince the maintainer to add his class
into your ifelse() chain. Failing this, he could just overwrite the
function \texttt{rss()}, with predictably disastrous results. In
reality, it's probably better to do all of these things with one method.
R's S3 methods mean that R developers can utilise a common interface
without having to update it when new classes come along. It also means
overloading clashes are less likely.

So let us create an S3 method to demonstrate.

First define the S3 method with \texttt{UseMethod()}

\begin{Schunk}
\begin{Sinput}
rss <- function(x) UseMethod("rss")
\end{Sinput}
\end{Schunk}

This creates the building block of an S3 method, the ``root'', if you
will.

Here we have specified that our method will be called \texttt{rss}. Now
we need to create the special cases of rss - the methods
\texttt{rss.rpart}, \texttt{rss.gbm}, and \texttt{rss.randomForest},
where the sections of code after \texttt{rss.} are the classes of object
we want them to work on.

\begin{verbatim}

rss.rpart <- function(x){
  
  sum((residuals(x)**2))
  
}

rss.gbm <- function(x){
  
  sum(x$residuals**2)
 
}

rss.randomForest <- function(x){

  res <- x$y - x$predicted  
  
  sum(res**2)  
    
}
\end{verbatim}

A default method can also be created - \texttt{rss.default} - which, as
the name suggests, is the default method when the argument \texttt{x} is
not a class that has a specific version of the method defined.

\begin{Schunk}
\begin{Sinput}
rss.default <- function(x, ...){
  
  warning(paste("RSS does not know how to handle object of class ", 
                class(x), 
                "and can only be used on classes rpart, gbm, and randomForest"))
          
          }
\end{Sinput}
\end{Schunk}

In this case a warning is issued, to let the user know that the object
class they were using was not appropriate.

We can now apply the \texttt{rss} method to an \texttt{rpart} model

\begin{Schunk}
\begin{Sinput}
rss(fit.rpart)
\end{Sinput}
\begin{Soutput}
#> [1] 10.17245
\end{Soutput}
\end{Schunk}

Also observe what happens when the object used is not of the decision
tree classes

\begin{Schunk}
\begin{Sinput}
rss(lm.fit)
\end{Sinput}
\begin{Soutput}
#> Warning in rss.default(lm.fit): RSS does not know how to handle object of
#> class function and can only be used on classes rpart, gbm, and randomForest
\end{Soutput}
\end{Schunk}

This guide to S3 methods was written to provide R users with the minimal
amount of information to start building their own S3 methods. For a more
complete treatment on S3 methods, see Advanced-R (Wickham 2014), R
Packages (Wickham, Hadley 2015), and the official S3 documentation (Team
2016, R Core Team (2012)).

\section*{References}\label{references}
\addcontentsline{toc}{section}{References}

\hypertarget{refs}{}
\hypertarget{ref-chambers1992}{}
Chambers, John M, and Trevor J Hastie. 1992. \emph{Statistical Models in
S}. CRC Press, Inc.

\hypertarget{ref-R}{}
R Core Team. 2012. \emph{R: A Language and Environment for Statistical
Computing}. Vienna, Austria: R Foundation for Statistical Computing.
\url{http://www.R-project.org/}.

\hypertarget{ref-rclassmethods}{}
Team, R Core. 2016. \emph{R: Class Methods}. R Foundation for
Statistical Computing.
\url{https://stat.ethz.ch/R-manual/R-devel/library/base/html/UseMethod.html}.

\hypertarget{ref-wickham2014}{}
Wickham, Hadley. 2014. \emph{Advanced R}. CRC Press.

\hypertarget{ref-wickham2015}{}
Wickham, Hadley. 2015. \emph{R Packages}. O'Reilly Media, Inc.

\address{%
Nicholas Tierney\\
Queensland University of Technology\\
Level 8, Y Block, Main Drive, QUT, Brisbane, Australia\\
}
\href{mailto:nicholas.tierney@gmail.com}{\nolinkurl{nicholas.tierney@gmail.com}}

\end{article}

\end{document}